\documentclass[epjCONF]{svjour}
\usepackage{graphics}
\usepackage[varg]{txfonts} 
\usepackage[latin1]{inputenc}
\session-title{HCP2011, Paris, France}
\begin{document}
\title{Triggering on electrons and photons with CMS}
\author{Alexandre Zabi\inst{1}\fnmsep\thanks{\email{Alexandre.Zabi@cern.ch}} }
\institute{Laboratoire Leprince Ringuet CNRS-IN2P3, Ecole polytechnique, 91128 Palaiseau CEDEX, France} 
\abstract{Throughout the year 2011, the Large Hadron Collider (LHC)
has operated with an instantaneous luminosity that has risen
continually to around $4\times10^{33} \rm{cm}^{-2}\rm{s}^{-1}$.  With
this prodigious high-energy proton collisions rate, efficient
triggering on electrons and photons has become a major challenge for
the LHC experiments.  The Compact Muon Solenoid (CMS) experiment
implements a sophisticated two-level online selection system that
achieves a rejection factor of nearly 10$^{6}$. The first level (L1)
is based on coarse information coming from the calorimeters and the
muon detectors while the High-Level Trigger (HLT) combines fine-grain
information from all sub-detectors. In this intense hadronic
environment, the L1 electron/photon trigger provides a powerful tool
to select interesting events. It is based upon information from the
Electromagnetic Calorimeter (ECAL), a high-resolution detector
comprising 75848 lead tungstate (PbWO$_4$) crystals in a ``barrel''
and two ``endcaps''. The performance as well as the optimization of
the electron/photon trigger are presented.
} 
\maketitle
\section{Introduction}
\label{intro}

The CMS detector has been designed to study the result of
proton-proton and heavy ion collisions produced by the LHC. These
experiments are conducted with the purpose of searching for new
particles and processes as well as revealing the very nature of the
elementary particle interactions~\cite{cite:jinst-cms}. From the
millions of collisions produced per second only 300 events per second
can be stored offline. Such a huge number of collisions is necessary
as these physics signatures are rare compare to the profusion of
QCD-induced background processes. The search for new physics crucially
relies on the trigger system performance that is used to select
them~\cite{cite:tdr-trigger-daq}.  The CMS trigger system is organised
in two consecutives steps~\cite{cite:trigger} : the Level-1 trigger
performs an event selection (custom-made electronics processors) based
on coarse energy deposits in the calorimeters and the muon systems
(output rate up to 100~kHz), followed by the HLT, implementing precise
selection algorithms (in commercial computers) based on finer
granularity and higher resolution information from all sub-detectors
in regions of interest identified at L1 (output rate about
300~Hz). The CMS ECAL provides a precise measurement of the energies
and positions of incident electrons and photons for both triggering
and offline analysis purposes. The energy measured by the hadronic
calorimeter (HCAL) is used to better identify and isolate
electromagnetic signals. A set of configuration parameters enables the
performance of the electron/photon trigger to be optimized for the
wide range of luminosities expected at the LHC.

\section{From ECAL to the Level-1 trigger} 
\label{sec:1:trigger}

\subsection{ECAL and the trigger primitive generation}

The CMS ECAL, composed of a Barrel (EB) and two Endcaps (EE),
comprises 75848 lead tungstate (PbWO$_4$) scintillating crystals
equipped with avalanche photodiode (APD) or vacuum phototriode (VPT)
light detectors in the EB and EE respectively. A Preshower detector
(ES), based on silicon sensors, is placed in front of the endcap
crystals to aid particle identification. The ECAL is highly segmented,
radiation tolerant and has a compact and hermetic structure, covering
the pseudorapidity range to $|\eta| < 3.0$.  Its target resolution
is 0.5\% for high-energy electrons/photons. It provides excellent
identification and energy measurements of electrons and photons, which
are crucial to searches for many new physics signatures. In the EB, 5
strips of 5 crystals (along the azimuthal direction) are combined into
trigger towers (TTs) corresponding to a 5$\times$5 array of
crystals. The arrangement in the EE is similar but more complicated
due to the X-Y layout of the crystals. The transverse energy ($E_T$)
detected by the crystals in a single TT is summed into a trigger
primitive (TP) by the front-end electronics and sent to off-detector
Trigger Concentrator Cards (TCCs) via optic fibres.

\subsection{Electron/Photon trigger path and algorithm}

\begin{figure}
\begin{center}
\resizebox{0.60\columnwidth}{!}{\includegraphics{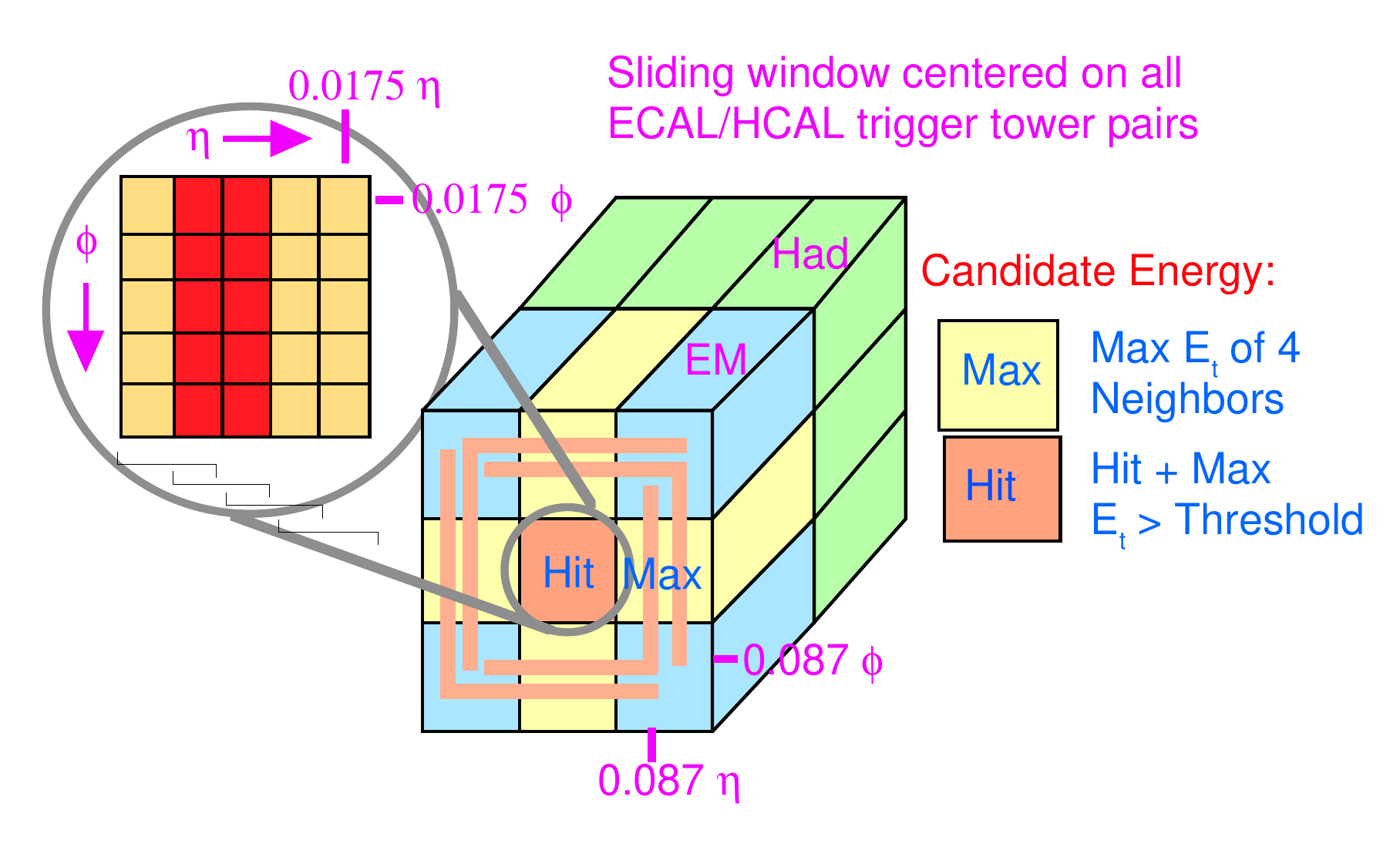}}
\caption{The Level-1 electron/photon trigger algorithm. The candidate
$E_T$ is the sum of the central TP (orange) and highest $E_T$ TT from
the 4 broadside neighbours (yellow). The Fine Grain profile (left box) and
the ratio with HCAL TT energy H/E (green) are used as vetoes while the
quiet corners (orange) are used to separate isolated from non-isolated
candidates. The HCAL TTs are aligned with the ECAL TTs in
pseudorapidity.}
\label{fig:l1algo} 
\end{center}
\end{figure}

The TCCs forward groups of TPs to the Regional Calorimeter Trigger
(RCT), which in turn combines pairs of TPs into L1 trigger candidates
in each region of interest ($4\times4$ TT). The Global Calorimeter
Trigger (GCT) then sends the four most energetic candidates to the
Global Trigger (GT), which generates the final L1 decision by applying
$E_T$ threshold cuts (named EG thresholds in the case of ECAL-based
candidates).  The electron/photon algorithm is based on a 3$\times$3
trigger tower sliding window as shown in Figure~\ref{fig:l1algo}. The
$E_T$ of an electron/photon candidate corresponds to the central TP of
the sliding window summed with the largest deposit in one of its 4
neighbour towers adjacent by side. Electromagnetic showers being
characterized by a compact lateral extension, candidates must have
their central tower containing 2 adjacent strips with a significant
fraction of the tower $E_T$ (typically 90\%). This criterion
characterized by 1 bit is called the Fine Grain (FG) veto bit that is
enabled only for candidates with $E_T >$~4~GeV. Moreover, the
associated HCAL energy contribution is required to be below a
threshold (typically H/E~$<$~5\% only for the central tower and for
candidates with $E_T >$~2~GeV). Non-isolated electron/photon
candidates require passing the previous criteria. In addition, the
isolated candidates must have a quiet neighbourhood characterized by
at least five adjacent TT among the 8 nearest ones with their
transverse energy below a threshold of 3.5~GeV.

\section{Online anomalous signals and their suppression}
\label{sec:2:spikes}

Anomalous signals were observed in the EB shortly after collisions
began in the LHC: these were identified as being due to direct
ionization within the APDs on single crystals, thus producing fake
isolated signals, with high apparent energy. These ``spikes'' can
induce large trigger rates at both L1 and HLT if not removed from the
trigger decision. On average, one spike with $E_T >$ 3~GeV is observed
per 370 minimum bias triggers in CMS at $\sqrt{s}$ = 7~TeV. If
untreated, 60\% of the EM trigger candidates, above an EG threshold of
12 GeV, would be caused by spikes. At high luminosity these would be
the dominant component of the 100~kHz CMS L1 trigger rate band
width~\cite{cite:spike-petyt}.

In the CMS ECAL the energy of an electromagnetic (EM) shower is
distributed over several crystals, with up to 80\% of the total energy
in a central crystal (where the electron/photon is incident) and most
of the remaining energy in the four adjacent crystals. This lateral
distribution can be used to discriminate spikes from EM signals.  A
``Swiss-cross'' topological variable $s=1-E_4/E_1$ ($E_1$ : $E_T$ of
the central crystal; $E_4$ : summed $E_T$ of the 4 adjacent crystals)
has been implemented offline to serve this purpose. A similar
topological variable has also been developed for the on-detector
electronics: the ``strip Fine Grain Veto Bit'' (sFGVB). Every TP has
an associated sFGVB that is set to 1 (signifying a true EM energy
deposit) if any of its 5 constituent strips has at least two crystals
with $E_T$ above a programmable ``sFGVB threshold'', of the order of a
few hundred MeV. If the sFGVB is set to zero, and the trigger tower
$E_T$ is greater than a ``killing threshold'', the energy deposition
is considered spike-like. The trigger tower energy is set to zero and
the tower will not contribute to the triggering of CMS for the
corresponding event. A detailed emulation of the full L1 chain has
been developed in order to optimize the two thresholds.

In order to determine the removal efficiency, data were taken without
the sFGVB or killing thresholds active. Spike signals identified
offline (with the ``Swiss cross'') were then matched to L1 candidates in
the corresponding RCT region and the emulator used to evaluate the
fraction of L1 candidates that would have been eliminated. In a
similar fashion the efficiency for triggering on real
electrons/photons could be estimated.

\begin{figure}
\begin{center}
\resizebox{0.53\columnwidth}{!}{\includegraphics{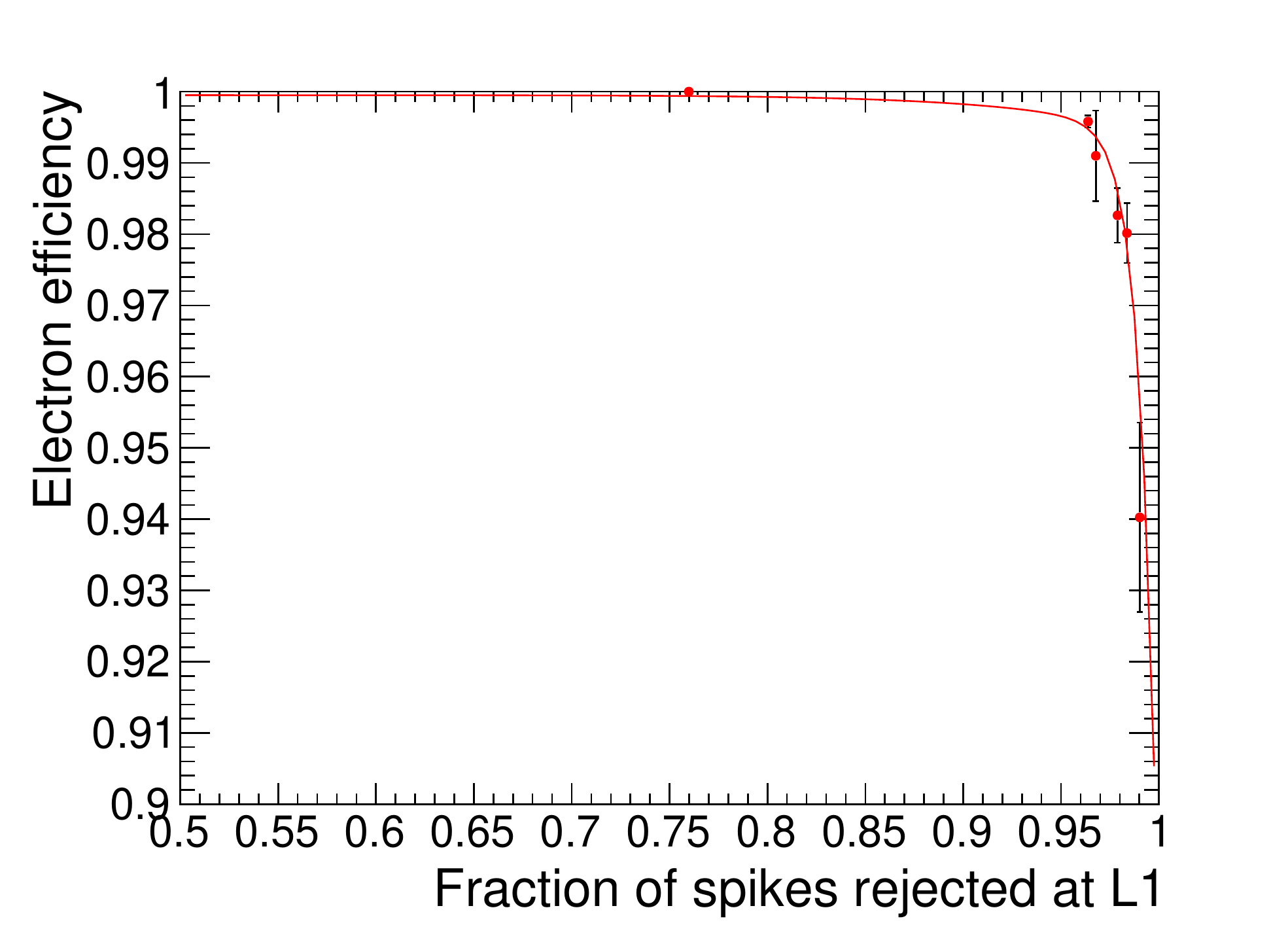} }
\resizebox{0.42\columnwidth}{!}{\includegraphics{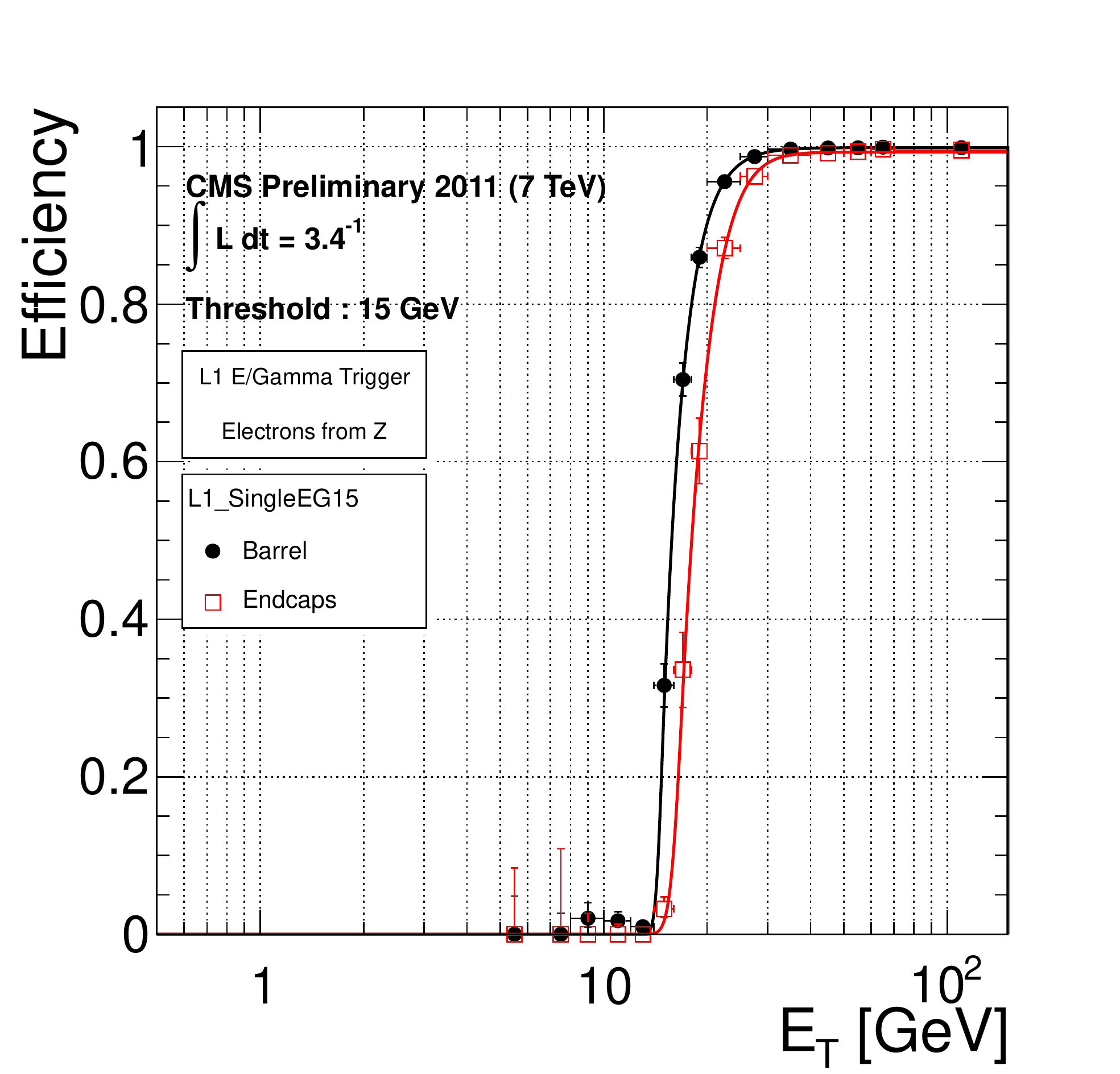} }
  \caption{(a) Electron efficiency as a function of the spike
rejection at L1 (spike removal ``sFGVB'' threshold set to 258 MeV;
``killing threshold'' set to 8 GeV). (b) Electron trigger efficiency
at L1 (``EG'' threshold : 15 GeV ET ), as a function of ET for
electrons in the ECAL Barrel (black dots) and Endcaps (red dots). An
unbinned likelihood fit was used.}
  \label{fig:perf}
\end{center}
\end{figure}

Three killing thresholds have been emulated ($E_T$ = 8, 12, and 18
GeV), combined with six sFGVB thresholds (152, 258, 289, 350, 456, 608
MeV). Figure~\ref{fig:perf} shows the electron efficiency (fraction of
electrons triggered after spike removal) versus the L1 spike rejection
fraction, for all sFGVB thresholds mentioned above (one point for each
threshold value) and a killing threshold of 8 GeV. The optimum
configuration was chosen to be an sFGVB threshold of 258 MeV and a
killing threshold of 8 GeV. This corresponds to a rejection of 96\% of
the spikes, whilst maintaining a trigger efficiency for electrons
above 98\%. With these thresholds the efficiency for higher energy
electrons is even larger: 99.6\% for electrons with $E_T >$
20~GeV. This optimized configuration was tested online at the
beginning of 2011. It gave a rate reduction factor of about 3 (for an
EG threshold of 12 GeV), and up to a factor of 10 for $E_T$ sum
triggers (which calculate the total EM energy in the whole calorimeter
system).

\section{Performance of the Level-1 electron/photon trigger}
\label{sec:3:trig-perf}

The trigger efficiency has been measured with electrons from
Z$\rightarrow$ee events, using a tag and probe method. The tag
electron is required to trigger the event at L1. The probe electron is
used for the efficiency studies. Both tag and probe electrons are
required to pass tight identification and isolation cuts. The
triggering efficiency is given by the fraction of probes which trigger
a given EG threshold, as a function of the probe $E_T$. In order to
trigger, the location of the highest energy trigger tower within the
electron supercluster must match a corresponding region of an L1
candidate in the RCT.

The trigger efficiency curves are shown in Figure~\ref{fig:perf} for
an EG threshold of 15 GeV.  The transverse energy on the x-axis is
obtained from the fully reconstructed offline energy. In the EE this
energy includes the preshower energy that is not available at L1.  As
a consequence the trigger efficiency turn-on point for the EE is
shifted to the right with respect to the EB. For both EB and EE,
corrections for crystal transparency changes are not currently
available at L1, which further affects the turn-on curve. The width of
the turn-on curves is partly determined by the coarse trigger
granularity, since only pairs of trigger towers are available for the
formation of L1 candidates, which leads to lower energy resolution at
L1. In the EE the material budget in front of the detector causes more
bremsstrahlung which, together with the more complex trigger tower
geometry in the EE, causes the turn-on curve to be wider than that for
the EB.  The main sources of inefficiency are caused by masked regions
(noisy or faulty : 0.2\% in the Barrel and 1.3\% in the Endcaps),
giving a plateau of 99.7\% in the EB and 98.8\% in the EE. The effect
on efficiency of the L1 spike removal has been verified to be
negligible, but this will require further optimization as the number
of collisions per bunch crossing increases in the future.

\section{Conclusion}

Over the course of 2011 the instantaneous luminosity provided by the
LHC has increased from about $10^{30}\rm{cm}^{-2}\rm{s}^{-1}$ to more
than $4\times 10^{33}\rm{cm}^{-2}\rm{s}^{-1}$. Optimizing the
electron/photon trigger performance, including the rejection of
spikes, has been a major challenge. A reprogramming of the front-end
electronics and ECAL TCC has allowed the implementation and
optimization of a spike killer at L1, which rejects a majority of
spikes ($>$96\%) whilst having a negligible impact on electron/photon
triggering efficiency. The results presented here display excellent
overall performance of the electron/photon trigger and demonstrate the
flexibility of this system.

\end{document}